\documentclass[12pt]{JHEP} 

\usepackage{epsfig}

\newcommand\fverb{\setbox\pippobox=\hbox\bgroup\verb}
\newcommand\fverbdo{\egroup\medskip\noindent%
			\fbox{\unhbox\pippobox}\ }
\newcommand\fverbit{\egroup\item[\fbox{\unhbox\pippobox}]}
\newbox\pippobox

\newcommand{\be}{\begin{equation}}
\newcommand{\ee}{\end{equation}}
\newcommand{\bea}{\begin{eqnarray}}
\newcommand{\eea}{\end{eqnarray}}
\newcommand{\ba}{\begin{array}}
\newcommand{\ea}{\end{array}}

\def\bbox{{\,\lower0.9pt\vbox{\hrule \hbox{\vrule height 0.2 cm
\hskip 0.2 cm \vrule height 0.2 cm}\hrule}\,}}
\newcommand{\dsl}{\pa \kern-0.5em /}

\newcommand{\nn}{\nonumber \\}
\catcode`@=11 \@addtoreset{equation}{section} \catcode `@=12

\def\* {&=&}
\def\del {\partial}
\def\eq#1{(\ref{#1})}
\def\zb{{\bar z}}
\def\mhat{{\hat M}}
\def\D{{\cal D}}
\def\Db{{\bar {\cal D}}}
\def\eps{\epsilon}

\title{ Nielsen-Olesen Vortices in
Noncommutative Abelian Higgs Model}

\author{by
Dileep P. Jatkar$^{a,b}$, Gautam Mandal$^{b,c}$ and Spenta R. Wadia$^c$\\
$^a$ {\it Mehta Research Institute of Mathematics
  and Mathematical Physics\\
Chhatnag Road, Jhusi, Allahabad 211 019, India}\\
\vspace*{1ex}

$^b$ {\it Theory Division, CERN \\
CH-1211, Geneva, 23, Switzerland}\\

\vspace*{1ex}

$^b$ {\it Department of Theoretical Physics, Tata Institute of
Fundamental Research\\
Homi Bhabha Road, Mumbai 400 005, India}
\\
E-mail: \email{dileep@mri.ernet.in, 
gautam.mandal@cern.ch, wadia@tifr.res.in}}

\preprint{\hepth{0007078}}

\abstract{
We construct Nielsen-Olesen vortex solution in the noncommutative
abelian Higgs model.  We derive the quantized topological flux of the
vortex solution. We find that the flux is integral by explicit
computation in the large $\theta$ limit as well as in the small
$\theta$ limit. In the context of a tachyon vortex on the
brane-antibrane system we demonstrate that it is this topological
charge that gives rise to the RR charge of the resulting BPS
D-brane. We also consider the left-right-symmetric gauge theory which
does not have a commutative limit and construct an exact vortex
solution in it.
}


\begin{document} 

\section{Introduction and Summary}

Noncommutative field theories appear naturally in the low energy
description of string theory in a constant Neveu-Schwarz
antisymmetric tensor background \cite{dougconnes}-\cite{nekra}.
They have also appeared previously in the study of $c=1$ matrix model
(two-dimensional string theory) \cite{DMW} and of two-dimensional QCD
\cite{DMW2}.  In \cite{doughull} it was observed that D-branes
compactified on a torus with constant Neveu-Schwarz $B$ field
background gives rise to an effective noncommutative field theory on
the compactified world-volume.  There has been some study of
perturbative dynamics of these theories\cite{seib1}. However, their
utility in understanding nonperturbative aspects of field theories has
attracted more attention lately\cite{gms}. Presence of constant
Neveu-Schwarz antisymmetric tensor field background tangential to the
D-brane leads to noncommutative gauge theory on the D-brane world
volume.  In case of the abelian gauge theory it has been shown that
the abelian noncommutative gauge theory is related by field
redefinition to the Born-Infeld electrodynamics action on the
D-brane\cite{SW}. Noncommutative gauge theories have been 
further discussed in \cite{spenta,grossnekra}; \cite{grossnekra}
discusses monopole solutions in noncommutative $U(1)$ gauge theory in
$3+1$ dimensions. 

The solitons in noncommutative scalar field theories studied in
\cite{gms} are non-topological, and are stable at large
$\theta$. These solutions lack stability in the commutative limit as
they violate Derrick's theorem. However, coupling of these scalars to,
say, gauge fields could add stability to these solutions.  Simplicity
of the noncommutative field theory formalism in the large $\theta$
limit was exploited in \cite{mukhi,harvey} to gain better
understanding of the tachyon condensation phenomenon in the non-BPS
D-branes\cite{sen} using the physics of noncommutative solitons.

It is important to generalize the class of noncommutative solitons to
accommodate solitonic solutions which have topological conserved
charge.  It is clear from the work of
\cite{mukhi,harvey,sen,witten,sen1}
that the topological solitons are relevant to the study of brane descent
relations in the non-BPS brane dynamics. It has been conjectured that
the topological charge of the tachyonic soliton on the brane
worldvolume of the non-BPS D$p$ brane is the RR charge of the
D$(p-k)$ brane\cite{sen,sen1}, where $k<p$ and for $k=1$ the soliton is a
kink, for $k=2$, it is a vortex and so on. Study of topology in
noncommutative field theories, however, is also important in its own right.

Here we will address this question in the abelian Higgs model. We will
show that the noncommutative abelian Higgs model supports topological
vortex solution with quantized flux, which can be any arbitrary
integer. The model that we will study first is a left module on the
Hilbert space, that is, in the complex Higgs field notation, the gauge
field multiplies the Higgs field $\phi$ from the left whereas for
$\bar\phi$ it multiplies from right. In section 2, we will briefly
review the Nielsen-Olesen vortex solution in the ordinary abelian Higgs
model, highlighting the first order,{\it i.e.}, Bogomolnyi formulation.

We generalise the Bogomolnyi equations to the noncommutative abelian
Higgs model in section 3 and  obtain exact solutions in the large
$\theta$ limit.  In this limit it is possible to do a systematic
$1/\theta$ expansion and we obtain corrections to the solution of the
leading order equations of motion.  We find that these corrections
converge quite rapidly. In the large distance limit this solution
matches with the large distance Nielsen-Olesen ansatz for vortex
solution in the ordinary abelian Higgs model. We also derive the
Seiberg-Witten(SW) map for the noncommutative abelian Higgs model. We
find that to order $\theta$, the SW map for the gauge field $A$ and
the gauge transformation parameter $\lambda$ is unaltered whereas the
SW map for the Higgs $\phi$ is linear in $\phi$. It has been argued in
the pure gauge theory case\cite{SW}, that the SW map
in the zero slope limit gives a field
redefinition from an ordinary field theory with the Born-Infeld action
to the noncommutative gauge theory action which is quadratic in the
gauge field strength. The SW map for $A$ is a nonlinear function of
$A$, whereas, as mentioned above, the SW map for $\phi$ is linear in
$\phi$. Therefore it is tempting to conjecture that the corresponding
action for the Higgs field in the ordinary field theory will retain
its form with each term in the Higgs Lagrangian multiplied by
some function of the gauge field $A$.

We also determine the profile of the magnetic field $B$ due to this
vortex solution. In the original coordinates it is just a
$\delta$-function. This is as expected since in the large $\theta$ limit,
the dispersion generated by the derivative terms in the kinetic
energy is totally suppressed. In the scaled coordinates, however, the
magnetic field profile is proportional to the ground state
wavefunction of the harmonic oscillator. The coefficient in front of
this wavefunction encodes the topological charge of the solution.

In section 4, we discuss the issue of the topology of this
solution. We show that this solution carries the topological conserved
charge (magnetic flux) which is determined by the behaviour of the
Higgs field. This topological charge is conserved, quantized and takes
integer values.  We establish this result in both small as well as 
large $\theta$ limit. In the small $\theta$ limit the leading result
reproduces the integer vortex charge of the commutative limit; we
show that higher order terms in $\theta$ expansion are total
derivatives and fall too rapidly at large distances to contribute to
the charge. In the large $\theta$ limit too, the leading term itself
gives the entire charge and subleading terms in the $1/\theta$
expansion do not contribute. We discuss the connection of this
topological charge with various other quantized charges and
indices. We also discuss the ``semiclassical'' limit of the field
configuration in which the topology reduces to that of the commutative
limit.

In section 5, we address the same question but for the left-right
symmetric module. In this case we get an interesting vortex solution
which at the face value seems to have charge 1 (using Witten's
identification \cite{witten} of the vortex charge with Atiyah-Singer
index of the Higgs field). The solution is not square integrable,
however; in the operator language the trace of the topological charge
over the one particle Hilbert space diverges and requires
regularisation. The question of finding a consistent regularisation
scheme to compute this charge is left for the future.

\section{Ordinary Abelian Higgs Model}

In this section we will briefly review the Nielsen-Olesen vortex solution to
the abelian Higgs model\cite{nielsen}, emphasising the Bogomolnyi limit of
this model\cite{bogo}. In the next section we will generalise these equations
to the noncommutative abelian Higgs model and work only in the Bogomolnyi
limit.

Let us start with the usual Abelian Higgs model. The Lagrangian is
\be\label{abelhiggs}
{\cal L} = -{1\over 4} F_{\mu\nu}F^{\mu\nu}-{1\over 2}|(\partial_{\mu}
+ieV_{\mu})\Psi|^2-\lambda(|\Psi|^2-|\Psi_0|^2)^2.
\ee

As mentioned earlier we will concentrate on the Bogomolnyi limit. 
Bogomolnyi limit and the corresponding first order equations of motion are
obtained from the energy functional by writing it in terms of strictly
positive quantities and a topological charge density and then minimising the
energy functional. We will take the static
ansatz. That means we will set $V_{0} = 0$ and 
$\partial_t V_m = 0 = \partial_t \phi$. Energy functional of the 
abelian Higgs model for such an ansatz is
\be
E = \int d^2x \left[ {1\over 4} F_{mn}F^{mn} + |D_m\Psi|^2
+\lambda(|\Psi|^2-\Psi_0^2)^2\right]
\ee
Let us rescale coordinates and fields and write them down in terms
of the following dimensionless variables
\bea
\Psi &=& \Psi_0\phi,\quad V_m = {\Psi_0}A_m,\quad 
x_m = {1\over e\Psi_0}y_m,\quad 
z={1\over\sqrt{2}}(y_1+iy_2)\equiv r\ e^{i\varphi} \quad{\rm and}\nn
E &=& 2\pi \Psi_0^2 {\cal E} 
\label{definitions}
\eea
The Nielsen-Olesen ansatz for the asymptotic form of a vortex solution
with winding number $n$ is
\bea
\phi \* \exp(i n\varphi)= {z^n\over (z\bar z)^{n\over 2}},
\quad\quad \bar\phi = \phi^* \nn
A_z \* -i{n\over 2z}, \quad\quad A_{\bar z} = A_z^*
\label{n-o-com}
\eea
The second equation is equivalent to 
\be
A_m\ dx^m = n\ d\varphi
\ee
With the definitions \eq{definitions} the energy
functional becomes
\be
{\cal E} = {1\over 2\pi} \int d^2z\left[{1\over 2}B^2
+ | \D_m \phi|^2 + {\beta\over 2}(\phi\bar\phi-1)^2
\right]
\label{energy-comm}
\ee
where $\beta = 2\lambda/e^2$, ${\cal D}$ is a covariant derivative
with  gauge field $A$ and
\be
B = \del_1 A_2 - \del_2 A_1 = -i(\del \bar A - \bar \del A)
= [\D, \Db],
\ee
It is easy to rewrite this as (see Appendix B
for the derivation in the noncommutative
case) 
\be
{\cal E} = {1\over 2\pi} \int d^2z\left[{1\over 2}(B+
(\phi\bar\phi-1))^2 + {\cal D}_{\bar z}\phi{\cal D}_z\bar\phi 
+ {\beta-1\over 2}(\phi\bar\phi-1)^2 + {\cal T} \right]
\label{energy}
\ee
where
\be
{\cal T} = \del_m S^m + B
\label{top-comm}
\ee
\be
S^m = {1\over 2} \epsilon^{mn}(i \phi {\cal D}_n\bar\phi-i
{\cal D}_n \phi \bar\phi) 
\label{def-sm-comm}
\ee
Our convention for $\eps^{mn}$ is that $\eps_{12}=1$.

We will argue below that ${\cal T}$ is a topological
density, which   generalises naturally to the noncommutative
case as well. It is easy to see in the commutative case 
(the same will be true for the noncommutative
generalisation) that 
${1\over 2\pi}\int d^2 z {\cal T}$  gives the magnetic flux of the vortex
and is an integer $n$. The energy functional for a vortex with winding 
number $n$, then, is
\be
{\cal E} = n + {1\over 2\pi}\int d^2z[{1\over 2}(B+(\phi\bar\phi-1))^2 +
{\cal D}_{\bar z}\phi{\cal D}_z\bar\phi+{\beta-1\over 2}(\phi\bar\phi-1)^2]
\label{energy2}
\ee
Notice ${\cal E}$ is a sum of absolute square terms except the last term.
However, when $\beta=1$, this term drops out and minimum of ${\cal E}$ can 
be obtained if the Bogomolnyi equations are satisfied:
\be
{\cal D}_{\bar z}\phi =0,\quad {\cal D}_z\bar\phi =0,\quad B = 
1- \phi\bar\phi 
\label{bogomolnyi}
\ee

It is interesting to note that the Euclidean action \eq{energy}
can be  written
in an elegant form using Quillen's superconnection
\cite{quillen} ${\cal A}$ defined below
\be
E=\int d^2 x \left( || {\cal F} -1 ||^2 + {\rm Str} \exp[{\cal F}]
\right)
\label{e-quillen}
\ee 
where in the exponential only the term proportional to the
volume form contributes, namely
\be
\int d^2x\, {\rm Str}\exp[{\cal F}] = F^+ + \bar{\cal D}\phi\wedge 
{\cal D}\bar\phi -\{ F^+ , \phi\bar\phi\}.
\ee
Here ${\cal F}$ is the curvature of the superconnection
\cite{quillen} ${\cal A}$
\be
{\cal A} = \left( \begin{array}{cc}   d + A^+  &  \phi \\
			\bar \phi &  d  \end{array}
\right)
\ee
and $F^+$ is the curvature of $A^+$.
The Bogomolnyi equations \eq{bogomolnyi} can be written
in a very suggestive form
\be
{\cal F} - 1=0
\label{b-quillen}
\ee

\section{Noncommutative Abelian Higgs Model}

Now let us consider noncommutative generalisation of these equations.
We present our conventions and definitions for noncommutative field theory 
following \cite{spenta} (see also appendix A). 
The energy functional for the noncommutative abelian Higgs 
model is given by \cite{spenta}
\be
{\cal E} = {\rm Tr}\left[{1\over 2}
B^2 +{\cal D}_{\bar z}\phi{\cal D}_z\bar\phi
+{\cal D}_z\phi{\cal D}_{\bar z}\bar\phi+
{1\over 2}(\phi\bar\phi-1)^2\right].
\label{nc-energy}
\ee
where $B$ is now defined as
\be
B = [\D,\Db]=
-i\left(\del_z A_{\bar z} - \del_{\bar z} A_z
\right) - [A_z, A_{\bar z}]
\label{def-b}
\ee
Following the  steps described in Appendix B, we can recast the energy 
functional as
\be
{\cal E} =  {\rm Tr}\left[{1\over 2}(B + (\phi\bar\phi-1))^2
+{\cal D}_{\bar z}\phi{\cal D}_z\bar\phi +{\cal T})\right],
\label{e-exp-nc}
\ee
where the topological term is now
\be
{\cal T}= {\cal D}_m S^m + B
\label{top-density-nc}
\ee
with $B$ defined as in \eq{def-b} and
\be
{\cal D}_m S^m = \del_m S^m - i[A_m, S^m]
\ee
$S^m$ is defined as before \eq{def-sm-comm} with
due attention to operator ordering now. The
use of the covariant derivative on $S^m$ 
in \eq{top-density-nc} is necessary since $S^m$
is now gauge-{\it covariant}.

We will argue in the next section that 
\be
I= {\rm Tr}\ {\cal T},
\label{top-nc}
\ee
corresponds to a topological charge. In 
particular,
for the noncommutative  Nielsen-Olesen
vortex of charge $n$ constructed below, $I$ evaluates to $n$ 
for any value of the noncommutativity parameter $\theta$.

\vspace{2ex}
\noindent{\it Bogomolnyi equations}
\vspace{2ex}

It is clear from \eq{e-exp-nc} that the  Bogomolnyi equations
remain the same as in \eq{bogomolnyi}, namely
\be
{\cal D}_{\bar z}\phi =0,\quad {\cal D}_z\bar\phi =0,\quad B = 
1- \phi\bar\phi 
\label{bogomolnyi-nc}
\ee
which are now to be interpreted as operator equations 
(conventions described in Appendix A).

It is difficult to solve the Bogomolnyi equations  exactly.
We will find the solution in various limits. First we will
solve the large $\theta$ limit ($\theta\to \infty$ and
$\frac{1}{\sqrt\theta}$ corrections). 

\subsection{Large $\theta$}
We will now consider the limit when the  the 
noncommutativity parameter $\theta$, defined by
\be
[X^1, X^2] = i\theta
\ee
is large. Let us define the following scaled operators
\be
X^i = \sqrt{\theta} \widetilde{X^i},\quad  i=1,2
\ee
We define the annihilation and creation operators as
\be
a= {1\over \sqrt 2} (\widetilde{X^1} + i\widetilde{X^2}),
\quad
a^\dagger = {1\over \sqrt 2} (\widetilde{X^1} - i\widetilde{X^2}) 
\ee
The scaled complex coordinates $w,\bar w$ are defined   as
\be
z = \sqrt{\theta}w,\quad \bar z = \sqrt{\theta}\bar w
\label{rescale-coordinates}
\ee
In accordance with the fact that the gauge potential
$A$ is a 1-form and the magnetic field is a 2-form,
the scaled  gauge potential $\tilde A$ and scaled
magnetic field $\tilde B$ are given by
\be
A = {\widetilde A\over \sqrt{\theta}},
\quad B = {\widetilde B\over \theta}.
\label{rescale}
\ee
With the above rescalings the energy functional \eq{e-exp-nc} becomes
\be
{\cal E} = \theta\ {\rm Tr}\left[{1\over 2}\left({
\widetilde B\over \theta} +
(\phi\bar\phi-1)\right)^2+
{1\over \theta}{\cal D}_{\bar w}\phi{\cal D}_w\bar\phi 
+ {\widetilde B\over\theta} +{\D_mS^m\over\theta}\right].
\label{e-exp-scaled}
\ee
Here $\del_w = - {\rm Ad}\, a^\dagger,
\del_{\bar w} = {\rm Ad}\, a$ (see Appendix A).

Let us now solve the Bogomolnyi equations
\eq{bogomolnyi-nc} in these rescaled variables
order by order in $1/\theta$. 

We write the following $1/\theta$-expansion of the Higgs field and the 
gauge field.
\bea
\phi \* \phi_\infty + \frac{1}{\theta} \phi_{-1} + \ldots \nn
A \* {1\over\sqrt{\theta}}\widetilde A=
{1\over\sqrt{\theta}} (A_\infty + \frac{1}{\theta}A_{-1} + \ldots)
\label{theta-exp}
\eea
The expansion of $\bar A$ is identical to that of $A$ except that $\bar A_i$
are hermitian conjugates of $A_i$. 
The large $\theta$ expansion of the magnetic field \eq{def-b} is
given by 
\be
B = {1\over \theta}\widetilde B
= {1\over \theta} B_\infty + {1\over \theta^2}B_{-1} + \ldots
\label{large-theta-b}
\ee
where
\be
B_\infty = -i(\del_w \bar A_\infty - 
\del_{\bar w} A_\infty)-[A_\infty,\bar A_\infty] 
= i([a^\dagger,\bar A_\infty] + [a, A_\infty])-[A_\infty, \bar A_\infty], 
\label{b0-exp}
\ee
\bea
B_{-1} \* -i(\del \bar A_{-1} - \bar\del A_{-1})-[A_\infty,\bar A_{-1}]-
[A_{-1},\bar A_\infty] 
\nn
\* i([a^\dagger, \bar A_{-1}]+[a,A_{-1}])-[A_\infty,\bar A_{-1}]-
[A_{-1},\bar A_\infty].
\label{b1-exp}
\eea

Let us substitute the expansions \eq{theta-exp}, \eq{large-theta-b} in
\eq{e-exp-scaled} or in the Bogomolnyi equations and solve these
equations order by order in $1/\theta$.

\vspace{2ex}
\noindent{\it $o(\theta)$  Bogomolnyi equations}
\vspace{2ex}

The relative orders of the Bogomolnyi equations are
easiest to figure out from \eq{e-exp-scaled}.
The $o(\theta)$ term in \eq{e-exp-scaled} gives
the leading Bogomolnyi equation
give 
\be
\phi_\infty \bar\phi_\infty =1.
\label{zero-order-phi}
\ee
This equation is solved by Witten\cite{witten} and the solution is
\be
\phi_\infty = {1\over\sqrt{a^na^{\dagger n}}}a^n,\qquad
\bar\phi_\infty = a^{\dagger n}{1\over\sqrt{a^n a^{\dagger n}}}.
\label{phi-sol}
\ee
It is easy to see by simple substitution that this ansatz indeed solves the 
leading order Bogomolnyi equation \eq{zero-order-phi}.

It is interesting to note a more general set of solutions of
\eq{zero-order-phi}, namely
\bea
\phi_\infty &=& {1\over\sqrt{(a+w_1)
\ldots (a+ w_n)\ (a^\dagger + \bar w_n)
\ldots (a^\dagger + \bar w_1)}}
\left[(a+w_1)
\ldots (a+ w_n)\right]
\nn
\bar \phi_\infty &=& \left(\phi_\infty \right)^\dagger
\label{phi-sol-general}
\eea
It is possible to find a solution for the gauge field
and the Higgs field in a $1/\theta$ expansion around this
more general solution, although we will not
explicitly write it down here. The interpretation
of this solution is simple for (a) all $w_i$
coincident and (b) far separated $w_i$. (a) represents
a vortex with all properties the same as \eq{phi-sol},
except that it is translated from the origin of the 
NC plane to the point $w_1$. (b) represents $n$ single 
vortices located
at $w_1,w_2,\ldots,w_n$. In the
case $\sum_i w_i=0$ the centre of
mass of the $n$ vortices is
fixed at $w=0$; the solution
\eq{phi-sol-general} then
describes the relative moduli
space of $n$ solitons. We end the
discussion of \eq{phi-sol-general} with a comment about
the index of ${\phi_\infty}$.  The kernel of $
{\phi_\infty}$ is given by
a linear span of the $n$ approximately orthonormal coherent
states $\{ |-w_1 \rangle, |-w_2 \rangle, \ldots |- w_n
\rangle \}$; this corresponds to index $
{\phi_\infty} = n$.

\vspace{2ex}
\noindent{\it $o(1)$  Bogomolnyi equations}
\vspace{2ex}

At $o(1)$ in \eq{e-exp-scaled} we get 
the Bogomolnyi equations involving the covariant 
derivative of the Higgs field, namely
\be
\del_{\bar w}\phi_\infty - i \bar A_\infty \phi_\infty =0,
\quad\quad 
\del_w\bar\phi_\infty + i\bar \phi_\infty A_\infty =0
\ee
For the sake of simplicity we will work only with the $\phi$ equation of 
motion ($\bar\phi$, being hermitian conjugate of $\phi$, can be determined 
from the solution to the $\phi$ equation of motion).

The $\phi$-equation can be written in the operator form as
\be
[a,\phi_\infty]=i\bar A_\infty \phi_\infty.
\label{sublead}
\ee
Let us recall at this point the action of creation and annihilation
operators of the harmonic oscillator on the one particle Hilbert space.
\be
a|m\rangle = \sqrt{m}|m-1\rangle,\qquad a^{\dagger}|m\rangle = 
\sqrt{m+1}|m+1\rangle .
\label{create}
\ee
For future purposes, it is useful to write down the action of the 
Higgs field on the one particle Hilbert space. The Higgs field
configuration for the vortex solution is written in terms of harmonic
oscillator creation and annihilation operators. Using \eq{create}, it is easy 
to see that the action of the Higgs field on the Hilbert space is
\be
\phi|m\rangle = |m-n\rangle .
\ee
Now let us look at the subleading equation, that is \eq{sublead}. Using the 
above results it is easy to derive the gauge field $\bar A_\infty$
($A_\infty$ is given by its hermitian conjugate). Thus 
\bea
\bar A_\infty \* -i \frac{1}{\sqrt{N+1}} a (\sqrt{N}-\sqrt{N+n})\nn
A_\infty \* i(\sqrt{N}-\sqrt{N+n})a^\dagger\frac{1}{\sqrt{N+1}},
\label{gauge-sol}
\eea
where $N= a^\dagger a$ is the number operator.

To gain a better understanding of the operator solutions \eq{phi-sol}
and \eq{gauge-sol} let us evaluate their expectation values
in a coherent state
\be
| w \rangle = e^{w a^\dagger}| 0 \rangle
\ee
By using the result 
\be
\langle w | f(a,a^\dagger) | w \rangle = \langle 0 |
f(a+w, a^\dagger + \bar w) | 0 \rangle
\ee
it is easy to see that in 
the large $w$ limit or equivalently in the large $\langle N\rangle$
limit, the expectation values become
\bea
\langle w | \phi_\infty | w \rangle \* \exp(in\varphi), \nn
\langle w | \bar A_\infty | w \rangle 
\* i \frac{n}{2 \bar w},\quad\quad 
\langle w | A_\infty | w \rangle = -i \frac{n}{2w}
\label{expectation}
\eea
The large distance behaviour here  is exactly the same as 
the large distance behaviour of the usual Nielsen-Olesen
vortex \eq{n-o-com}%
\footnote{In stead of calculating expectation
values in coherent states one could alternatively
evaluate  the Weyl-Moyal (inverse) map of the operator
solutions to find the classical functions on the NC plane;
the large distance behaviour  obtained
this way is the same as that in \eq{expectation}
or \eq{n-o-com}.}.

In a way, this result is expected because in the large $\theta$ limit
we have ignored the derivative terms and then in the large $\langle N\rangle$
limit the asymptotic behaviour of the vortex solution becomes exact. Behaviour
of the vortex solution in the finite domain of the $w$ plane, that is,
for finite values of $\langle N\rangle$ depends on the competition between the
kinetic energy terms and the potential energy terms in the energy functional. 
Exact solution to the equations in the large $\theta$ expansion, which
essentially ignores the kinetic energy effect, of the Bogomolnyi equations,
reduces to the potential energy minimisation in the large $\langle N\rangle$
limit. We will show below that \eq{top-nc} evaluates to $n$ for this solution. 
Therefore, this solution carries topological charge, which is determined by
the quantized magnetic flux through the vortex solution.

\vspace{2ex}
\noindent {\it The magnetic field}
\vspace{2ex}

{}From  \eq{b0-exp} we see that $B_\infty$ is given entirely
in terms of $A_\infty$. Using the solution  \eq{gauge-sol} 
in \eq{b0-exp} (details in Appendix C) we get 
\be
B_\infty = n |0\rangle\langle 0| ,
\label{b-not}
\ee
where $|0\rangle\langle 0|$ is a projection operator onto
the vacuum state. It is curious that the leading term in the large $\theta$
expansion of the magnetic field has such a remarkably simple form. It is also
interesting to note that the trace of $B_\infty$ in the one particle Hilbert 
space
gives us exactly the integer $n$ as desired. We will elaborate more on this
when we will discuss the topology of our solution. In terms of the original
unscaled coordinates this vacuum projection operator is essentially a
$\delta$-function. However, in the scaled variables the vacuum projection
operator is represented by the ground state wavefunction, i.e., by the
Gaussian. Significance of this will be discussed in the next section.
 
\vspace{2ex}
\noindent{\it $o(1/\theta)$ and higher orders}
\vspace{2ex}

So far we have looked at the leading and the first subleading term in
the large $\theta$ expansion of the Bogomolnyi equations. Here we will
look at the higher corrections to the Higgs field $\phi$ as well as
the gauge field $A$. It is interesting to see that the large $\theta$
correction are quite small and the convergence of the solution is
remarkably fast.

The $o(1/\theta)$ equation of motion is given by
\be
B_\infty = -\phi_\infty\bar\phi_{-1}-\phi_{-1}\bar\phi_\infty.
\label{bnot-eq}
\ee
Notice that since we have already determined $B_\infty$,
we can use the above equation to solve for  $\phi_{-1}$. We get
\be
\phi_{-1} = -{n\over 2}|0\rangle\langle n|,\qquad\qquad \bar\phi_{-1} = 
-{n\over 2}|n\rangle\langle 0|.
\label{phi-one}
\ee
As mentioned in the beginning of this subsection the solution has a very good
convergence property. To see this it is instructive to write the leading order
solution \eq{phi-sol} as follows:
\be
\phi_\infty = \sum_{m=0}^{\infty} |m\rangle\langle m+n|,\quad {\rm and}
\quad \bar\phi_\infty = \sum_{m=0}^{\infty} |m+n \rangle\langle m| .
\ee
Thus the leading order solution involves an infinite sum whereas the first
correction contains only one term as can be seen from \eq{phi-one}.

The subleading correction to the gauge field solution is
obtained by solving the $o(\theta^{-2})$ equation
\be
\bar\del\phi_{-1} - i\bar A_\infty\phi_{-1} -i\bar A_{-1}\phi_\infty =0
\ee
The solution is
\be
\bar A_{-1} = -i{n\over 2}\sqrt{n+1}|0\rangle\langle 1| .
\ee
Substituting this result in \eq{b1-exp} we can determine first subleading 
correction to the magnetic field 
\be
B_{-1} = n(n+1)\left(|1\rangle\langle 1| - |0\rangle\langle 0| \right)
\label{sub-B}
\ee
Note that the correction has a vanishing trace, ensuring that
\be
{\rm Tr}( B_\infty + \frac{1}{\theta} B_{-1}) = {\rm Tr} B_\infty =n
\ee 
\subsection{Finite $\theta$}

Having constructed the vortex solution for large $\theta$, we now ask
what happens to this solution at finite $\theta$. To do this let us
look at the Seiberg-Witten(SW) map for the noncommutative abelian Higgs
system. 

Recall the SW map for the pure Yang-Mills theory is
\bea
\hat S(A)\equiv \hat A_i(A) \* A_i -{1\over 4}\theta^{kl}
\{ A_k,\partial_lA_i+F_{li}\} +{\cal O}(\theta^2)\nn
\hat S(\lambda)\equiv \hat\lambda(\lambda, A) \* \lambda + 
{1\over 4}\theta^{kl}\{ \partial_k\lambda,A_l\} +{\cal O}(\theta^2).
\label{gauge-sw}
\eea
We wish to carry out this exercise for $\phi$, {\it i.e.}, we look for 
a map $\hat S$
\be
(\phi, A) \mapsto \hat S(\phi, A) \equiv (\hat \phi, \hat A)
\ee
such that
\be
\hat S(\phi + i\lambda \phi, A + d\lambda)
= \left(\hat \phi + i\hat \lambda \phi, \hat A + d\hat \lambda +
i \left(\hat \lambda \hat  A - \hat A  \hat \lambda  \right)\right)
\ee
where both  the map $\hat S$ and the noncommutative
gauge transformation parameter $\hat \lambda$ are to
be found so as to satisfy the above equation. It turns out that
the map for $\hat A$ and $\hat \lambda$ given in \eq{gauge-sw} is unaltered. 
The map for $\phi$ is
\be
\hat \phi = \phi - \frac{1}{2} \theta \epsilon^{kl} A_k \del_l \phi
+ o(\theta^2).
\label{map-phi}
\ee
The SW map can be used for determining the change in the fields 
at any value $\theta=\theta_0$ due to small increment $\delta\theta$. At
$\theta=0$, right hand side of \eq{gauge-sw} and \eq{map-phi} contain ordinary
products of the fields, but if \eq{gauge-sw} and \eq{map-phi} are used for
determining a small increment at $\theta=\theta_0$ then right hand side of
these equations contain $\star$-products. 

There is an important difference between the SW map for the Higgs 
field $\phi$ and that for the gauge field $A$. Whereas the SW map 
for the gauge field $A$ is nonlinear in $A$ that for the Higgs field 
$\phi$ is linear in $\phi$. It is the nonlinearity of the SW map for the gauge
field which was useful\cite{SW} in relating the noncommutative gauge
theory action to the Born-Infeld action for the ordinary gauge theory.
It is easy to see by 
successive transformations that the SW map for $\phi$ remains linear in $\phi$
but becomes a nonlinear function of $A$. Therefore, it is tempting to 
conjecture that the Higgs action in terms of the ordinary Higgs field would
still be of the same form, albeit multiplied by complicated functions of the
gauge field $A$.

By using the above result, it is easy to see that the
equations of motion, written in terms of the ordinary commutative 
fields, remain the same at $r \to \infty$ as in
the usual abelian Higgs model case. Therefore, the asymptotic
form of the vortex solution, written above, is
valid. Though the  solution in the noncommutative problem will
differ from the abelian case in the bulk, for our purposes
here, especially in the next section, where we discuss the topology
of the vortex solution, detailed form of the solution
in the bulk is not relevant.\footnote{
A numerical vortex solution has been constructed in
\cite{schaposnik} in the context of a somewhat
different action.}

\section{Topology}

It is interesting to ask what happens to topology
when one studies noncommutative gauge theories coupled to
matter. Below we show that  the topological charge $I$
in \eq{top-nc}  is independent of $\theta$ and 
therefore the configuration space of noncommutative theory
splits into  the same topological sectors as
in the commutative case. We first show this
for small $\theta$.

\subsection{Small $\theta$ expansion}

We rewrite \eq{top-nc} in the Moyal form
(by that we mean using ordinary
functions and star products). Thus,
\be
I= \frac{1}{2\pi} \int d^2 z (\D_m S^m + B)
\label{top-nc-moyal}
\ee
where the covariant derivatives and the magnetic
field $B$ are defined, in the
Moyal formalism, in \eq{star-examples} in
Appendix A.

Besides the explicit $\theta$-dependence involved in the star product,
$S^m$ and $B$ also have an expansion in terms of $\theta$ since they
are built out of $\phi $ and $A_i$ which are solutions of the
Bogomolnyi equations (we imagine writing these
equations here in the Moyal form,
therefore explicitly containing $\theta$, and solving them iteratively
in small $\theta$). We write
\bea
\phi \* \phi_0 + \theta \phi_1 + \ldots
\nn
A_i \*  A_{i,0} + \theta A_{i,1} + \ldots
\eea
The small $\theta$ expansion for  $B= \epsilon^{kl}(\del_k A_l -i A_k
\star A_l)$ and the topological charge \eq{top-nc-moyal}
are
\be
B = B_0 + \theta B_1 + \ldots
\ee
\be
I = I_0 + \theta I_1 + \ldots
\ee
The zero-order term evaluates to
\be
I_0= {1\over 2\pi}\int d^2 x B_0 =
{1\over 2\pi}\oint A_{i,0} dx^i  = n
\ee
This follows from \eq{n-o-com} since in the
zero-th order in $\theta$ the Bogomolnyi
equations are identical to those of  the ordinary
abelian Higgs model. We have  also used the fact 
that  at zero-th order 
\be
\int d^2 x\ \left[\D_m S^m \right]_0 = \oint r\ d\varphi  
\ S_{r,0}|_{r=\infty} =0
\ee
as can be seen explicitly from the asymptotic
form \eq{n-o-com}.

We now carry out the iterative solution to
first nontrivial order in $\theta$. Thus,
for example,
\be
B_1 = {1\over 2} \epsilon^{kl}\epsilon^{ij}\del_k A_{i,0} \del_l A_{j,0}
        + \epsilon^{kl}\del_k A_{l,1} 
    =   -   \left({i\over 2} \epsilon^{kl}\del_k\phi_0 \del_l \bar\phi_0 
        +  \phi_0 \bar \phi_1 + \phi_1\bar \phi_0\right)
\label{small-B}
\ee
It is easy to see that the contribution
of the magnetic field to $I_1$ becomes
a total derivative. The
expression after the first equality is
given by
\bea
B_1&=& \del_k f^k
\nn
f^k &\equiv& \eps^{kl}
\left( A_{l,1} + {1\over 2} \eps^{ij} A_{i,0} \del_l A_{j,0}
\right)  
\eea
The contribution of $B_1$ to the topological charge
at this order, therefore, is
\be
I_1 = \int d^2 x\ B_1 = 
\oint r\ d\varphi\  f_r |_{r=\infty} =0
\ee
It is easy to see that $f_r$ vanishes as $f_r \sim 1/r^3$.
Similar arguments can be made to show that the
contribution of $\D_m S^m$ becomes a total
derivative too, and the surface term at
$r=\infty$ vanishes. 

These arguments, namely that (a) the integrals involved in $I_n$ all
become surface terms, and (b) the surface terms vanish at $r=\infty$,
in fact generalize to all higher orders in $\theta$. The proof of (a)
is s straight-forward generalisation of the first-order calculation;
regarding (b) we need to only use the fact, proved at the end of the
previous section that higher $\phi_n, A_{i,n}$ have a faster fall-off
at $r=\infty$ than the zero order solutions.  Regarding the rate of
fall-off of the surface terms in successive $I_n$ it is easy to see
from a scaling argument (using scaled coordinates
$(1/\sqrt\theta)x^i$) that order $\theta^n$ terms are down by $2n$
powers of derivative, or equivalently by $1/r^{2n}$ in the large $r$
limit.

To summarise,  we see that the higher order terms in the
small $\theta$ expansion of $I$ all vanish and therefore do not
modify the topological charge.

\subsection{Large $\theta$  expansion}

We now show that the topological charge \eq{top-nc} is the integer $n$
also in the large $\theta$ limit and would like to argue that the
successive orders in the $1/\theta$ expansion do not modify this
result.

{}From the expression  \eq{b-not}, namely
\be
B_\infty = n |0\rangle\langle 0| ,
\ee
it is easy to see that
\be
{\rm Tr}\ B_\infty =n.
\label{tr-b0}
\ee
We emphasize here that to see the result \eq{tr-b0}, one must consider
the subleading $1/\theta$ terms as in the last section.  In the
$\theta=\infty$ limit, as indicated in \eq{large-theta-b}, $B=0$. In
other words, since ``Tr'' in \eq{tr-b0} is actually $\theta \int d^2
z$ in the original coordinates, it is essential to keep the $1/\theta$
terms in $B$.

By using calculations similar to Appendix C, it is easy to
see that 
\be
{\rm Tr}\ \D_m S^m =0
\ee
using  the leading order expressions in large $\theta$. 

Thus, if one writes
\be
I = I_\infty + \theta^{-1} I_{-1} + \ldots
\ee
we have
\be
I_\infty = n.
\ee

It is easy to check, using results from the last
section, e.g. \eq{sub-B}, that 
\be
I_{-1}=0 .
\ee

Although we have not checked it in explicit detail, it
is easy to argue, in
keeping with our result for small $\theta$,
that the higher order terms in the large $\theta$ expansion
will all vanish too. This implies 
\be
I= n
\ee
nonperturbatively. 

\subsection{Comments}

We would like to make several comments:

\begin{enumerate}

\item Although we have ostensibly shown that $I$ evaluates to an
integer by using the on-shell vortex solution, it is clear from our
small $\theta$ arguments that the expression $I$ will evaluate to an
integer for all {\bf off-shell} configurations which satisfy the
conditions \eq{n-o-com} at $r=\infty$. 

\item  The topological charge $I$ we have introduced
can be identified with RR charge in the context of vortex solutions on
brane-antibrane systems. This fact has been shown by vertex operator
calculations in \cite{kennedy}. This provides additional evidence the
quantity $I$ we have introduced must be
quantized.

\item It has been remarked
in \cite{witten} that the topological
charge of the vortex is actually the
same as the index of $\phi$:
\be
I= \iota(\phi)= {\rm dim\ ker}\ \phi - {\rm dim\ coker}\ \phi
\label{witten-index}
\ee
Note that such a relation automatically 
implies a ``quantization'' (integer-valued-ness) of the topological
charge $I$. It is important to appreciate that
the ``quantization'' here does not refer to the usual quantization
(specified by a finite $\hbar$) but rather to
noncommutative field theory. In a commutative
field theory, with or without $\hbar$, there is no notion of a kernel of
$\phi$. It is a characteristic feature of
NCFT's that one can have ''quantization'' conditions
in a classical field theory.

\item One can see in a ``semiclassical''  sense  
how $I$ is also related
to the Atiyah-Singer index of the Dirac operator in the 
gauge field background of the vortex. We skip
the details here, but the main point is that if
one considers the  normal mode of a fermion field 
$\xi_m = z^m$ and adiabatically turns on the background
gauge field representing our vortex solution, then the normal mode
gets transformed to
\be z^m \mapsto \ \exp{i\!\int\!\! A}\cdot z^m = z^{m-n}.
\ee
By the standard arguments relating to spectral flow
and the index, one can see that the index of the 
Dirac operator is $n$ since the ``Fermi sea''
shifts by $n$. 

\item It would be very interesting to figure out why
our topological charge $I$ is equal \eq{witten-index} to the index of
$\phi$. It is natural, {\it e.g} from the
viewpoint of  the superconnection \cite{quillen}, 
that the Dirac equation for the fermion should be considered in the
background of both the gauge field and the Higgs field. In 
other words, the fermion zero modes should satisfy, schematically
\be
{\cal D}\!\!\!/ \xi \equiv (\del \!\!\!/ + A\!\!\! / + \phi) \xi =0
\ee
It is possible that the toplogical charge $I$ actually measures the
index of ${\cal D}\!\!\!/$. In that case, in the large
$\theta$ limit, $\del \!\!\!/ + A\!\!\! /$ will pick
up a factor of $1/\sqrt\theta$ and decouple, leaving just
the index of $\phi$ \footnote{We thank Mike Douglas for
a discussion on this point.}.

\item The zero-th order equation for $\phi$
in the large $\theta$ limit, namely  \eq{zero-order-phi}, is
the equation of a {\bf fuzzy circle} in the
configuration space. The calculation in
Appendix C can be interpreted to
mean, on the other hand, that the topological
charge $I$ receives contribution from states with
expectation values
\be
\langle a^\dagger a \rangle
= M \Rightarrow \langle (X^1)^2 + (X^2)^2\rangle =
2 \theta (M + \frac{1}{2})
\label{large-coordinate}
\ee 
where $M \to \infty$. Equation \eq{large-coordinate} can be 
regarded as a fuzzy circle in ``coordinate space''
\footnote{%
It is interesting to note the appearance of the zero
point energy in the expression for radius of the fuzzy
circle.}.
Thus, $I$ appears to characterise maps from
a fuzzy circle to a fuzzy circle.      
\end{enumerate}

We would also like to make a few remarks about the qualitative nature
of the vortex solution we have found.  Let us, in particular, discuss how
some features of the magnetic field \eq{b-not},
\eq{sub-B} in the large
$\theta$ expansion could be arrived at by the following physical
reasoning.

It can be seen from the form of $A_\infty$ and $\bar A_\infty$ in
\eq{gauge-sol} and the expression of $B_\infty$ in \eq{b0-exp} that
magnetic field has to involve equal number of creation and
annihilation operators.

Now, for the $n$ vortex solution, $\phi$ annihilates the ground state
as well as $n-1$ excited states in the one particle Hilbert
space. Topology of the solution is obtained by taking trace of
$B_\infty$ over the whole Hilbert space.  This can also be done as
proposed by Witten\cite{witten} by determining the index of
$\phi$. The index of $\phi$ is $n$ since the lowest $n$ states in the
Hilbert space are in the kernel of $\phi$. So we would expect that
$B_\infty$ can be written in terms of a linear combination of the
projection operator $|i\rangle\langle i|$, where $i = 0 \ldots n-1$.

As is well known (see, {\it e.g.} \cite{gms,grossnekra}), these
projection operators are represented in the Moyal form (in the sense
of Appendix A) in terms of Laguerre functions, related to harmonic
oscillator wavefunctions. In particular, $|i\rangle\langle i|$ is
represented by the $i$-th excited state wavefunction.  All these
wavefunctions have nodes except the ground state wavefunction. If the
magnetic field is written in terms of a particular $|i\rangle\langle
i|$ for some $i\not= 0$ then the magnetic field develops a zero in a
finite region in the $w$ plane and as a consequence of the Bogomolnyi
equations, the Higgs field relaxes to its vacuum value. This
configuration is allowed only at asymptotic infinity in the $w$
plane. One can, therefore, rule out the possibility that the magnetic
field has a form $|i\rangle\langle i|$ for some $i\not= 0$ since it
contradicts with the minimal energy ansatz of the Bogomolnyi
limit. One can also rule out the the possibility that 
this choice of $B$ 
corresponds to  $n$ single vortex solutions centred at different
locations in $w$ plane (such a solution is briefly mentioned
after \eq{phi-sol}). The reason is that the
$i$-th excited state wavefunction changes sign after encountering a
node, whereas change in the sign of the magnetic field would
correspond to an anti-vortex configuration. It is well known that a
vortex-antivortex configuration does not satisfy Bogomolnyi
condition. This still leaves the
possibility, however, that the magnetic field is a
linear combination of $|i\rangle\langle i|$ for $i$ belonging to a
certain subset of the Hilbert space in such a way that the resulting
functional form of $B$ never develops a zero. Profile of the magnetic
field of the vortex that we seem to get at the leading order indicates
that we have a single vortex of vorticity $n$ sitting at the origin. A
careful look at the subleading corrections to $B$ in the large
$\theta$ limit, however, shows that the general profile of the
magnetic field does allow for the linear combinations of
$|i\rangle\langle i|$ for $i\not= 0$. As has been shown at the end of
previous subsection these corrections to the magnetic field do not
affect the value of the topological charge.

\section{Higgs coupling to Diagonal U(1)}

So far we have considered the NC Higgs model which has a nontrivial
action only of the left $U(1)$.  Here we will consider a model which has 
left-right action. In other words, in NC abelian Higgs model, the Higgs field
couples to left and right abelian gauge fields which are different fields in
general, {\it i.e.}, NC abelian Higgs model has $U(1)_L\times U(1)_R$ gauge
symmetry. In case of left module, $U(1)_R$ gauge field is set to zero. In this 
section we will consider a model where left and right U(1) gauge fields are 
identified with each other.  This is of interest because this model
does not have a commutative analogue. For another recent discussion
of this model, see \cite{polychronakos}. 

It is easy to generalize \eq{bogomolnyi} to this
case and they are given by
\be
D_{\bar z}\phi =0,\quad D_z\bar\phi =0,\quad B = 
1-\phi\bar\phi+\bar\phi\phi,
\ee
where the covariant derivative is defined as
\be
D_{\bar z}\phi= \del_{\bar z} \phi - i A_{\bar z} \phi + i \phi A_{\bar z} 
\ee
and similarly for $\bar\phi$. These equations can be rewritten in terms
of the one form gauge field ${\cal A}$ as
\be
[{\cal A}, \bar{\cal A}] = \phi\bar\phi+\bar\phi\phi
\ee
where, $[{\cal A}, \bar{\cal A}]= 1-B$ \cite{gms}.

This equation supports exact solution unlike the left module NC abelian 
Higgs model. To arrive at this solution let us notice the following fact
about the commutator $[\chi^2,\psi]$, where $\chi$ and $\psi$ are any two 
operators.
\be
[\chi^2, \psi] = \chi [\chi,\psi] + [\chi,\psi] \chi 
= \{ \chi, [\chi,\psi]\}.
\ee
This relation can be used to convert the anticommutator on the right hand
side of the Bogomolnyi equation into a commutator. For our purpose, we choose
$\chi = \xi^2$, $\psi = \bar\xi^2$, and $\phi = 2\xi$ and
$\bar\phi = [\xi, \bar\xi^2]$. Substituting this in the
Bogomolnyi equations, it is easy to see that these equations are solved 
by 
\be
{\cal A} = \sqrt{2}\xi^2,\qquad \bar {\cal A} = \sqrt{2}\bar\xi^2.
\ee
Putting this into the second Bogomolnyi equation gives us the following 
relations
\be
[\xi^2, \xi] =0,\qquad [\bar\xi^2, [\xi, \bar\xi^2]]=0.
\ee
Of these equations, first one is trivially satisfied whereas the second
equation is also satisfies if
\be
[\xi, \bar\xi^2] = f(\bar\xi).
\ee
In particular, for the choice of $f(\bar\xi) = 2\bar\xi$ the
equation reduces to 
\be
[\xi, \bar\xi] =1
\label{a-adagger}
\ee
Thus the Bogomolnyi equations are solved by
\be
{\cal A} = \sqrt{2} \xi^2,\quad \bar{\cal A} = \sqrt{2}\bar\xi^2,
\quad \phi = 2\xi,\quad \bar\phi = 2\bar\xi. 
\ee

An example of solution to \eq{a-adagger} is
\be
\xi= a,\quad \bar\xi = a^\dagger .
\ee
Such a solution is obviously not normalizable. However
using arguments similar to that in \cite{witten} we see
that this solution has nontrivial winding number (=1),
because, firstly (1)  $\xi \sim r \exp(i\varphi)$
in the semiclassical limit and
secondly, (2) dim ker $\xi-$  dim ker $\bar\xi = 1 -0 = 1$.
It would be interesting to get this winding number from
a suitably regularised topological charge. 

\vspace{5ex}

\acknowledgments 

One of us (S.R.W.) would like to thank Luis Alvarez-Gaume and Avinash
Dhar for discussions in the initial stages of this work.  D.P.J. would
like to thank Amol Dighe for discussion and Theory Division of CERN
for hospitality. G.M. would like to thank Mike Douglas for a
discussion regarding the topological index.

\appendix
\section{Noncommutative Field Theory: operator conventions}

We collect below the conventions and some results that
are needed for the purposes of our paper.
For more details, see \cite{spenta}.

\vspace{2ex}
\noindent{\it The Weyl-Moyal map}:
\vspace{2ex}

We consider a one-particle Hilbert space
${\cal H}$ carrying a representation
of the Heisenberg algebra
\be
[X^1, X^2] =i
\label{heisen}
\ee
For $f$ a function on $R^2$
\be
f(x^1,x^2) = {1\over 2\pi} \int dk_1\ dk_2
\  \tilde f(k_1,k_2) \exp[i k_1x^1+
i k_2x^2]
\ee
we define an operator $\mhat(f)$ on ${\cal H}$  
\be
f \mapsto \mhat(f)
\ee
by the rule
\be
\mhat(f) = 
{1\over 2\pi} \int dk_1\ dk_2
\  \tilde f(k_1,k_2) \exp[i k_1 X^1+
i k_2 X^2]
\ee
where $X^1, X^2$ are the operators in \eq{heisen}.
It is easy to see that
\bea
{\rm Tr} \mhat (f) &=& {1\over 2\pi} \int d^2x f(x^1,x^2)
\nn
\mhat (f) \mhat (g) &=& \mhat (f \star g)
\nn
\mhat ({\del f\over \del x^i}) \* ( i\theta)^{-1} \epsilon_{ik}
[ X^k, \mhat(f)] 
\eea
where
\be
f \star g (x^1,x^2)=\left( \exp[{i\over 2}\theta 
\epsilon^{kl} {\del\over \del x^k}
{\del\over \del y^l}]\ f(x^1,x^2)\ g(y^1,y^2)\right)|_{y=x}
\ee
Clearly, NCFT can be defined either in terms of operators on ${\cal
H}$ or on ordinary functions whose multiplication is defined in the
sense of star product. We use the operator approach in most of
our paper. Thus, for
example, the equation \eq{nc-energy} is written in the operator
language; its alternative form (which we will call the
Moyal form) will be
\be
{\cal E} = {1\over 2\pi} \int
d^2 z \left[{1\over 2}
B \star B +{\cal D}_{\bar z}\phi
\star {\cal D}_z \bar\phi
+{\cal D}_z\phi \star {\cal D}_{\bar z}\bar\phi+
{1\over 2}(\phi \star  \bar\phi-1)\star (\phi \star  \bar\phi-1)
\right].
\label{nc-energy-moyal}
\ee
where
\bea
D_i \phi \* \del_i \phi - i A_i \star \phi
\nn
B \* \del \bar A - \bar \del A - i (A \star \bar A -
\bar A \star A)
\label{star-examples}
\eea

\noindent{\it Conventions for the noncommutative $U(1) \times U(1)$
gauge theory:}
\vspace{2ex}

Note that for NC gauge theory, $g \phi \not = \phi g$ even when $g$ is
(an operator representative of) a $U(1)$ transformation. Thus it is
important to distinguish between the left $U(1)$ from the right
$U(1)$.

We define the left- and right- gauge fields as
$A_i$ and $A'_i$ and the
corresponding gauge transformations as
$\delta_\lambda$ and $\delta_{\lambda'}$ respectively.
Our conventions for the left $U(1)$ are
\vspace{3ex}
\bea
\delta_\lambda A_i \* \del_i \lambda + i [\lambda,A_i]
\nn
\delta_\lambda \phi \* i \lambda \phi
\nn
\delta_\lambda {\bar \phi} \* -i {\bar \phi} \lambda
\eea
Similarly for the right $U(1)$
\bea
\delta_{\lambda'} A'_i \* \del_i \lambda' + i [\lambda',A'_i]
\nn
\delta_{\lambda'} \phi \*  - i \phi\lambda' 
\nn
\delta_{\lambda'} {\bar \phi} \* i  \lambda' {\bar \phi}
\eea
The covariant derivatives, when both the gauge fields are
non-zero, are given by 
\bea
D_i \phi = \del_i \phi - i A_i \phi + i \phi A'_i
\nn
D_i {\bar\phi} = \del_i {\bar \phi} + i\bar \phi A_i
 - i A'_i \bar \phi
\eea

In the above formulae we mean the derivatives and
products as in the operator formulation explained
above.

In Sections 2,3 and 4, we consider only the left $U(1)$, by putting
$A'_i=0= \lambda'$. This has a 
commutative limit (the same as the right $U(1)$).
In Section 5, we put $A_i = A'_i$ (the gauge
transformations $\lambda = \kappa = \lambda'$) 
which corresponds to the
diagonal $U(1)$. Clearly the diagonal $U(1)$ does not
have a commutative counterpart since the relevant gauge
transformation, for example of $\phi$, is
\be
\delta_\kappa \phi = [\delta_\lambda \phi + \delta_{\lambda'}
\phi]_{\lambda=\lambda'=\kappa} = i[\kappa,\phi]
\ee
which disappears in the commutative limit.

\section{Derivation of the topological density}

In order to derive \eq{e-exp-nc} we need to show that
\be
(D_m S^m + B) + 2 \bar D\phi D\bar\phi + \frac{1}{2}
(B + (\phi \bar\phi -1))^2
= D\phi \bar D\bar \phi + \bar D\phi D\bar\phi 
+\frac{1}{2} B^2 +  \frac{1}{2} (\phi \bar \phi -1)^2
\ee
This is equivalent to
\be
D_m S^m = D\phi \bar D\bar \phi - \bar D\phi D\bar\phi
- \frac{1}{2} (\phi\bar \phi B + B \phi \bar\phi)
\label{eval-Dm-sm}
\ee
We start by observing that
\bea
D_z S^z \*  \frac{1}{2}\left(\del (\phi \bar D\bar \phi -
\bar D\phi\bar \phi) - i[A, \phi \bar D\bar \phi -
\bar D\phi\bar \phi]\right) \nn
{} \* \frac{1}{2}\left(
D\phi \bar D\bar \phi +
\phi D(\bar D\bar \phi) - D(\bar D\bar \phi).\bar \phi
- \bar D\phi D\bar\phi
\right)
\eea
Similarly
\be
D_\zb S^\zb = -\frac{1}{2}\left(\bar D\phi D\bar\phi +
\phi \bar D(D\bar \phi) - \bar D( D\bar \phi).\bar \phi
-D\phi \bar D\bar \phi 
\right)
\ee
Using these, and the fact that $[D, \bar D]\phi=B\phi,
[D, \bar D]\bar \phi= -\bar \phi B$ we get
\eq{eval-Dm-sm}.

\section{Calculation of $B_\infty$}

We give some steps in the evaluation of \eq{b0-exp}.

We give details how to calculate traces of  quantities like
\be
t \equiv [a, C]
\label{t-def}
\ee
for some operator $C$. Here $a$ is the usual annihilation
operator, satisfying
\be
a | m \rangle = \sqrt{m} | m-1 \rangle\quad \forall m=1,..
\ee  
Formally,
\bea
{\rm Tr }\ t \* \sum_{m=0}^M  \langle m | a\ C - C\ a | m \rangle
\nn
\* \sum_{m=0}^M \sqrt{m+1} \langle m+1 | C |m\rangle
- \sum_{m=1}^M \sqrt{m} \langle m | C |m-1\rangle
\label{t-sum}
\eea
where eventually we should take the limit $M\to \infty$ (this
regulator has been used in \cite{grossnekra}). There
are two ways in which  we can proceed from here:

Method 1: Shifting the summation variable in the second term,
we get
\be
{\rm Tr}\ t = \sqrt{M+1} \langle M+1 | C |M\rangle
\label{stokes}
\ee
The calculation from this viewpoint reduces to the
fuzzy circle at the boundary $\langle a^\dagger a
\rangle = M \to \infty$. This
actually provides a noncommutative version of
Stokes' theorem (recall that Ad$\ a$ actually
plays the role of a derivative operator).

Method 2: In this method, we note that the first
summation has a contribution from the ground state
which is missing from the second summation. We take care
of this fact by writing
\be C\ a | m \rangle = \sqrt{m} C (1 - 
|0 \rangle \langle 0|) | m-1 \rangle
\ee
Of course, the answers in both the methods are the
same.

The first two terms in \eq{b0-exp} are already of the form
\eq{t-def} (or hermitian conjugate). The two
methods described above also work for the third term
in a similar fashion. 

According to Method 1, the trace of \eq{b0-exp}
gets evaluated as
\be
{\rm Tr}\  B_\infty = {\rm Lim}_{M=\infty} 
2  \sqrt{M}(\sqrt{M+n} - \sqrt{M}) = n
\ee
The contributions come entirely from the first two terms
in \eq{b0-exp}.

According to Method 2, the first and second terms in
\eq{b0-exp} give
\be
2\left[(N - \sqrt{N(N+n)})(1 - |0 \rangle \langle 0|) -
(N+1 - \sqrt{(N+1)(N+n+1)}) \right]
\ee
whereas the third term gives
\be
-(\sqrt{N} - \sqrt{N+n})^2(1 - |0 \rangle \langle 0|) +
(\sqrt{N+1} - \sqrt{N+n+1})^2
\ee
Combining, we get
\be
B_\infty = n |0 \rangle \langle 0|,
\ee
namely equation \eq{b-not}.

\end{document}